\documentclass[english,pra,amsmath]{revtex4}
\usepackage[T1]{fontenc}
\usepackage[latin9]{inputenc}
\usepackage{amsmath}
\usepackage{graphicx}
\usepackage{amssymb}

\begin{document}

\author{William J. Mullin$^{a}$ and Asaad R. Sakhel$^{b}$}

\title{Generalized Bose-Einstein Condensation}

\affiliation{$^{a}$Department of Physics, University of Massachusetts, Amherst,
Massachusetts 01003 USA \\
 $^{b}$Al-Balqa Applied University, Faculty of Engineering Technology,
Basic Sciences Department, Amman 11134, JORDAN}

\date{\today}
\begin{abstract}
Generalized Bose-Einstein condensation (GBEC) involves condensates
appearing simultaneously in multiple states. We review examples of
the three types in an ideal Bose gas with different geometries. In
Type I there is a discrete number of quantum states each having macroscopic
occupation; Type II has condensation into a continuous band of states,
with each state having macroscopic occupation; in Type III each state
is microscopically occupied while the entire condensate band is macroscopically
occupied. We begin by discussing Type I or {}``normal'' BEC into
a single state for an isotropic harmonic oscillator potential. Other
geometries and external potentials are then considered: the {}``channel''
potential (harmonic in one dimension and hard-wall in the other),
which displays Type II, the {}``cigar trap'' (anisotropic harmonic
potential), and the {}``Casimir prism'' (an elongated box), the
latter two having Type III condensations. General box geometries are
considered in an appendix. We particularly focus on the cigar trap,
which Van Druten and Ketterle first showed had a two-step condensation:
a GBEC into a band of states at a temperature $T_{c}$ and another
{}``one-dimensional'' transition at a lower temperature $T_{1}$
into the ground state. In a thermodynamic limit in which the ratio
of the dimensions of the anisotropic harmonic trap is kept fixed,
$T_{1}$ merges with the upper transition, which then becomes a normal
BEC. However, in the thermodynamic limit of Beau and Zagrebnov, in
which the ratio of the boundary lengths increases exponentially, $T_{1}$
becomes fixed at the temperature of a true Type I phase transition.
The effects of interactions on GBEC are discussed and we show that
there is evidence that Type III condensation may have been observed
in the cigar trap. 
\end{abstract}
\maketitle

\section{Introduction\label{sec:Introduction}}

Bose-Einstein condensation (BEC) has been intensely researched in
recent years since the advent of laser and magnetic cooling of trapped
alkaline atoms achieved that state \cite{Ketterle1,Stringari,Leggett,PethickSmith}.
In normal Bose-Einstein condensation (NBEC) a macroscopic number of
particles populates a single quantum state (usually the ground state
of the system) below a critical temperature. The state is often the
lowest momentum state in a homogeneous system \cite{Einstein,London,Ziff} (for
an extensive review see Ref.~\cite{Ziff}), 
or it could be the lowest harmonic oscillator state in a trapped gas.
Many authors have considered alternative possibilities: for example,
a number of states may simultaneously be macroscopically occupied,
or a band of states is macroscopically occupied while no single state
has a macroscopic number of particles. Either of these last two cases
is called a {}``generalized Bose-Einstein condensation'' (GBEC).
The terminology {}``generalized'' was first used by Girardeau \cite{Girar1,Girar2,Girar3}
in considering a homogeneous interacting gas. (See also the work of
Luban \cite{Luban}). An early description of the phenomenon was by
Casimir \cite{Casimir} in treating a uniform gas in an elongated
box (the {}``Casimir prism''). Rehr and Mermin \cite{RehrMermin}
found that a rotating Bose gas had a GBEC. Other terminology has referred
to {}``smeared'' \cite{Girar1} and {}``fragmented'' condensations
\cite{Nozieres1,Nozieres2}. The most thorough analyses of GBEC have
been done by Van den Berg and coworkers \cite{VDB80,VDB81,VDB82,VDB82A,Pule,VDB83,VDB84,VDB86,VDB86A}
and more recently by Zagrebnov et al \cite{Zag98-1Int,Zag98-2Int,Zag98-3Int,Zag2000Int,Zag2000-1Int,Zag2004,Zag08Int,BeauZag,Beau}.
Ho and Yip \cite{Ho} claim that the spin-1 Bose gas is an example
of fragmentation.

However, there seem to be several schools of research on this topic
who do not know of and do not quote the others' work. For example,
Van Druten and Ketterle \cite{VDK} and more recently Beau and Zagrebnov
\cite{BeauZag} have theoretically discussed an example of GBEC. Ref.~\cite{BeauZag} 
has followed on the extensive work of the Van den
Berg group and does not quote Ref.~\cite{VDK}. On the other hand,
Ref.~\cite{VDK}, and many references in the literature to this paper,
do not refer to the previous papers of the Van den Berg or Zagrebnov
schools. A frequently quoted paper by Nozieres \cite{Nozieres2} presents
a proof to show that fragmented BEC is ruled out for repulsive interacting
systems; however, the proof holds only for one kind of GBEC; no proof
or reference appears for the other kinds of GBEC. Thus we believe
a pedagogic review of the subject is needed, to clarify the subject,
make the details more widely known, and perhaps to stimulate experimental
research to find cases of the phenomenon. 

Van den Berg et al \cite{VDB82A} have classified three general forms
of BEC: There is condensation into

I) a discrete number of quantum states each having macroscopic occupation
(of order $N$, the number of particles)---e.g., NBEC has a single
condensed state;

II) a band of states with each state having macroscopic occupation.

III) a band of states each state having only microscopic occupation
(the occupation number $n_{i}$ of each such state has $n_{i}/N\rightarrow0$
in the thermodynamic limit), but with the entire band having a macroscopic
number of particles.

We will present examples of each of these situations in the present
paper. All of our examples will be of ideal gases. The question then
arises whether a GBEC can be maintained in the case of interacting
systems. Noziéres \cite{Nozieres2} showed that interactions would
favor NBEC when there were repulsive interactions. Nevertheless there
are several cases in the literature where interacting cases of GBEC
have been given; we will discuss these later and how they avoid violating
the Noziéres analysis.

The treatments of Refs.~\cite{BeauZag} and \cite{VDK} are interesting
in that they bear on BEC in ultra-cold gases; they have theoretically
discussed a case of two-step GBEC with a first transition into the
lowest band of states in a {}``cigar''-shaped harmonic trap followed
at a lower temperature by a condensation into the lowest single-particle
state. The latter transition, into a single state, occurred at a transition
temperature $\sim1/\ln(N)$, which would seem to disappear in the
thermodynamic limit (TL). Each pair of authors considers an alternative
TL in which the lower transition \emph{persists} for large particle
number. We will discuss these cases more fully below.  Sonin \cite{Sonin}
was the first to note that there could be multiple BEC transitions in
a parallelepiped geometry corresponding to a variety of way of taking
the TL, although he did not discuss any associated GBEC. The original
case of a GBEC associated with the double BEC transition was considered
by Van den Berg and coworkers in a flat-plate geometry \cite{VDB83,VDB86}. There have been many
experiments with gases in cigar traps; we comment on their relevance
to the possible observation of GBEC.

We begin by discussing the normal BEC (NBEC) for an isotropic harmonic
potential followed by a sample of Type II condensation. Then we give
a detailed discussion of the anisotropic harmonic {}``cigar'' trap,
for which transitions are possible at \emph{two} different temperatures.
As discussed above the nature of the lower transition depends crucially
on how the TL is taken. Such a result shows a secondary purpose of
our paper: to show there is more than one way to go to the large-scale
limit; the density of states will depend on the relative values of
the respective dimensions of an ansiotropic container.  We treat the
``Casimir prism'' to illustrate
GBEC with box boundary conditions in all directions. The effects of
interactions on GBEC and the implications of recent experiments are
discussed in Sec.~\ref{sec:interactions}. A more abstract
analysis of the three kinds of GBEC is given in the Appendix.

\section{The isotropic 3D oscillator\label{sec:Normal-BEC--Isotropic-3D}}

To illustrate NBEC, we use the three-dimensional (3D) isotropic oscillator
potential rather than the usual 3D homogeneous ideal gas, because
it can show a very interesting GBEC when the potential is made anisotropic.
The harmonic potential is taken to be \begin{equation}
V=\frac{1}{2}\frac{U}{L^{2}}\left(x^{2}+y^{2}+z^{2}\right),\label{eq:Hamiltonian}\end{equation}
 where $U$ is the potential strength. Note that the harmonic potential
has been written so it has an evident length scale $L$ and the frequency
is \begin{equation}
\omega=\sqrt{\frac{U}{m}}\frac{1}{L},\label{eq:omega}\end{equation}
 where $m$ is the particle mass. We insert this length since taking
the thermodynamic limit in a harmonic potential involves increasing
the number of particles while \emph{weakening} the potential to maintain
constant overall density $\rho$. Having this length scale is also
necessary to see the relative sizes of the occupation numbers of various
states \cite{Damle ,WJM1D3D}. The harmonic length $a_{0}=\sqrt{\hbar/(m\omega)}$
is a measure of the rms deviation of a particle from the center of
the well and is not the appropriate density length scale as we will
see. The energy levels are\begin{equation}
\varepsilon_{p_{x}p_{y}p_{z}}=\hbar\omega\left(p_{x}+p_{y}+p_{z}+\frac{3}{2}\right),\end{equation}
 with $p_{i}$ taken over all positive integers.

The total number of particles in the system is given in terms of the
usual Bose distribution function as\begin{equation}
N=\sum_{p_{x},p_{y},p_{z}}\left\{ \exp\left[\beta\left(\hbar\omega\left(p_{x}+p_{y}+p_{z}\right)-\mu\right)\right]-1\right\} ^{-1},\end{equation}
 where $\beta=1/(k_{B}T)$ with $T$ the temperature, $k_{B}$ the
Boltzmann constant, and $\mu$ the chemical potential. We will always
consider the ground-state energy to be incorporated into the chemical
potential $\mu$ so that the ground state is effectively zero. We
introduce a temperature parameter \begin{equation}
T_{0}\equiv\frac{\hbar}{k_{B}}\sqrt{\frac{U}{m}}\frac{1}{a},\label{eq:T0}\end{equation}
 where $a=1/\rho^{1/3}$ is an average interparticle separation.

For high temperature $T$ we can replace the sums by integrals:\begin{eqnarray}
N & = & \int_{0}^{\infty}dp_{x}\int_{0}^{\infty}dp_{y}\int_{0}^{\infty}dp_{z}\frac{1}{\exp\left[\frac{T_{0}a}{TL}\left(p_{x}+p_{y}+p_{z}\right)-\beta\mu\right]-1}\nonumber \\
 & = & \frac{L^{3}}{a^{3}}\left(\frac{T}{T_{0}}\right)^{3}\int_{0}^{\infty}du_{1}\int_{0}^{\infty}du_{2}\int_{0}^{\infty}du_{3}\frac{1}{e^{\left(u_{1}+u_{2}+u_{3}+\alpha\right)}-1},\end{eqnarray}
 where $u_{i}=T_{0}ap_{i}/TL$ and\begin{eqnarray}
\alpha & = & -\beta\mu.\end{eqnarray}
The integral is most easily done by expanding the integrand in inverse
powers of the exponential to get\begin{eqnarray}
N & = & \frac{L^{3}}{a^{3}}\left(\frac{T}{T_{0}}\right)^{3}\int_{0}^{\infty}du_{1}\int_{0}^{\infty}du_{2}\int_{0}^{\infty}du_{3}\sum_{l=1}^{\infty}e^{-l\left(u_{1}+u_{2}+u_{3}+\alpha\right)}\nonumber \\
 & = & \frac{L^{3}}{a^{3}}\left(\frac{T}{T_{0}}\right)^{3}F_{3}(\alpha),\label{eq:noncondensed}\end{eqnarray}
 where the Bose integral \cite{Robinson} is \begin{equation}
F_{n}(\alpha)\equiv\sum_{l=1}^{\infty}\frac{e^{-l\alpha}}{l^{n}}.\label{eq:Ffunction}\end{equation}
 We see that the way to define a density parameter (or any thermodynamic
variable) so that it is scale independent \cite{Damle ,WJM1D3D} is
in terms of a volume $L^{3}$ with the parameter given by \begin{equation}
\rho=\frac{1}{a^{3}}=\frac{N}{L^{3}}.\end{equation}
 As the temperature decreases, Eq.~(\ref{eq:noncondensed}) can be
satisfied by decreasing $\alpha$ until the quantity $F_{3}(\alpha)$
has a maximum, $F_{3}(0)=\zeta(3)=1.202$ where $\zeta(n)$ is the
Riemann zeta function. The temperature corresponding to this maximum
is the critical temperature $T_{c}$ given by\begin{equation}
T_{c}=\frac{T_{0}}{\zeta(3)^{1/3}}.\end{equation}

Below the transition temperature the non-condensed particle number,
$N-N_{0},$ with $N_{0}$ the condensate number, is still given by
the right side of Eq.~(\ref{eq:noncondensed}) with $\alpha=0:$\begin{eqnarray}
N-N_{0} & = & \frac{L^{3}}{a^{3}}\left(\frac{T}{T_{0}}\right)^{3}\zeta(3)\nonumber \\
 & = & N\left(\frac{T}{T_{c}}\right)^{3},\end{eqnarray}
 from which we see that\begin{equation}
N_{0}=N\left[1-\left(\frac{T}{T_{c}}\right)^{3}\right].\end{equation}
 Assume that all the condensed particles fall into the ground state
(as usual); the occupation number corresponding to the ground state
is \begin{eqnarray}
n_{000} & = & \frac{1}{e^{\alpha}-1}\nonumber \\
 & \approx & \frac{1}{\alpha}=N_{0}=O\left(\frac{L^{3}}{a^{3}}\right),\end{eqnarray}
 where $O(x)$ means order of magnitude of $x$. Thus $\alpha$ is
small, but not actually zero. The low excited states have occupation
\begin{eqnarray}
n_{p_{x}p_{y}p_{z}} & = & \left\{ \exp\left[\frac{T_{0}a}{TL}(p_{x}+p_{y}+p_{z})+\alpha\right]-1\right\} ^{-1}\nonumber \\
 & \approx & \left[\frac{T_{0}a}{TL}(p_{x}+p_{y}+p_{z})+\alpha\right]^{-1}.\label{eq:npxpypz}\end{eqnarray}
 Since $\alpha$ is so small we can drop that term in Eq.~(\ref{eq:npxpypz})
as long as some $p_{i}>0$. In that case the occupation number is
$O(L/a)=O(N^{1/3}),$ which is small relative to the ground-state
occupation, $O(N)$. Only the ground state is macroscopically occupied.
All this is quite standard except here we have states of an oscillator
potential rather than free particles in a box. The BEC considered
here is Type I because only one state has macroscopic occupation.

Normal BEC might also include cases where there are a discrete number
of states each having macroscopic occupation. For example, the well-know
experiment \cite{MITInterf} in which two condensates were brought
together to form an interference pattern is such an example. One might
also consider multiple condensates each having a different spin orientation
\cite{Ho,LMSpinLetter}. Condensates trapped in multiple wells are
sometimes said to be {}``fragmented'' \cite{Alon}.

\section{The channel potential\label{sec:The-Canal}}

In a Type II condensation we have a band of states each with macroscopic
occupation. Here we discuss a rather peculiar case \cite{VDB81} in
which a spinless two-dimensional (2D) ideal gas has a harmonic potential
in the $z$ direction and is free in the $x$ direction with periodic
boundary conditions in that direction over a large distance $L$.
The potential forms what might be called a channel or trough. The
energy levels are\begin{equation}
\varepsilon_{k_{x}p_{z}}=\frac{\hbar^{2}k_{x}^{2}}{2m}+\hbar\omega\left(p_{z}+\frac{1}{2}\right),\end{equation}
 with the harmonic frequency given by Eq.~(\ref{eq:omega}); we use
the same $L$ parameter in both momentum and harmonic dimensions.
The total number of particles in the system is given in terms of the
usual Bose distribution function as\begin{equation}
N=\sum_{s_{x},p_{z}}\left\{ \exp\left[\beta\left(\frac{h^{2}s_{x}^{2}}{2mL^{2}}+\hbar\sqrt{\frac{U}{m}}\frac{p_{z}}{L}-\mu\right)\right]-1\right\} ^{-1}\end{equation}
 with $s_{x}=0,\pm1,\pm2\cdots$ . We define two temperature parameters\begin{equation}
T_{x}=\frac{h^{2}}{k_{B}2ma^{2}}\end{equation}
 and\begin{equation}
T_{z}=\frac{\hbar}{k_{B}}\sqrt{\frac{U}{m}}\frac{1}{a},\end{equation}
where the density is $\rho=1/a^{2}$. These temperatures are simply natural units
expressed in terms of the parameters of the problem. For simplicity
we assume the two parameters are the same: $T_{x}=T_{z}\equiv T_{0}$.
For high temperature $T$ we can change the sums to integrals:\begin{eqnarray}
N & = & \int_{-\infty}^{\infty}ds_{x}\int_{0}^{\infty}dp_{z}\frac{1}{\exp\left[\frac{T_{0}a^{2}}{TL^{2}}s_{x}^{2}+\frac{T_{0}a}{TL}p_{z}-\beta\mu\right]-1}\nonumber \\
 & = & \frac{L^{2}}{a^{2}}\left(\frac{T}{T_{0}}\right)^{3/2}\int_{-\infty}^{\infty}du\int_{0}^{\infty}dw\frac{1}{e^{\left(u^{2}+w+\alpha\right)}-1}\nonumber \\
 & = & N\left(\frac{T}{T_{0}}\right)^{3/2}\sqrt{\pi}F_{3/2}(\alpha),\label{eq:Nsum}\end{eqnarray}
since the average density is $\rho=N/L^{2}$. $F_{3/2}(\alpha)$ has
a maximum $\zeta(3/2)=2.61$ (as in the standard 3D homogeneous gas).
The critical temperature is \begin{equation}
T_{c}=T_{0}\left(\frac{1}{\sqrt{\pi}\zeta(3/2)}\right)^{2/3}.\end{equation}
 For $T<T_{c}$ there is a condensate whose number $N_{0}$ is given
by \begin{eqnarray}
N-N_{0} & = & N\left(\frac{T}{T_{0}}\right)^{3/2}\sqrt{\pi}F_{3/2}(0)\nonumber \\
 & = & N\left(\frac{T}{T_{c}}\right)^{3/2},\end{eqnarray}
 or the condensate fraction $f_{0}=N_{0}/N$ is\begin{equation}
f_{0}=\left[1-\left(\frac{T}{T_{c}}\right)^{3/2}\right].\label{eq:fzero}\end{equation}

\begin{figure}[h]
\centering \includegraphics[width=3in]{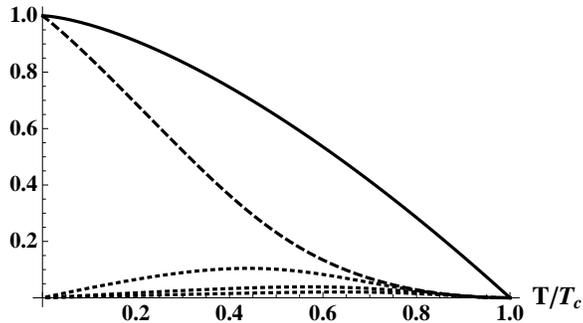}

\caption{The condensate distributions for the channel potential versus temperature.
The solid line is the entire condensate distribution $f_{0};$ the
dashed line is the that of the lowest momentum state $f_{0,0}=n_{0,0}/N$
of the condensate band; the dotted lines are the distributions of
the next higher momentum levels. }

\label{figChannel} 
\end{figure}

Assume that the ground state (that is, lowest $k-$state of the lowest
harmonic level $p_{z}=0$) is macroscopically occupied with number\begin{eqnarray}
n_{00} & = & \frac{1}{e^{\alpha}-1}\nonumber \\
 & \approx & \frac{1}{\alpha}=O\left(\frac{L^{2}}{a^{2}}\right).\end{eqnarray}
However, now the excited states \emph{cannot} be neglected as in Sec.
\ref{sec:Normal-BEC--Isotropic-3D}. Consider the entire band of $k$
states corresponding to the lowest harmonic level $p_{z}=0$. The
low excited states have occupation\begin{equation}
n_{s_{x},0}\approx\left[\beta\left(\frac{h^{2}s_{x}^{2}}{2mL^{2}}\right)+\frac{a^{2}\gamma}{L^{2}}\right]^{-1},\end{equation}
where we have written $\gamma=\alpha L^{2}/a^{2}$. Obviously $n_{s_{x},0}=O(L^{2}/a^{2})$
for \emph{all} $k$. On the other hand, we can easily see that the
harmonic bands of states with $p_{z}>0$ do \emph{not} contribute
macroscopically, being of order $L/a\ll N.$ We must sum up the contribution
$N_{b}$ of the first harmonic band ($p_{z}=0)$. We then have for
the band number \begin{eqnarray}
N_{0}=N_{b} & \equiv & \sum_{s_{x}=-\infty}^{\infty}\frac{1}{\exp\left[\frac{1}{N}\left(\frac{T_{0}}{T}s_{x}^{2}+\gamma\right)\right]-1}\nonumber \\
 & \cong & \sum_{s_{x}=-\infty}^{\infty}\frac{1}{\left[\frac{1}{N}\left(\frac{T_{0}}{T}s_{x}^{2}+\gamma\right)\right]}\nonumber \\
 & = & N\pi\sqrt{\frac{T}{T_{0}\gamma}}\coth\left(\pi\sqrt{\frac{T\gamma}{T_{0}}}\right).\label{eq:N0integral}\end{eqnarray}
 The sum in the last line is exact; replacing the sum in the first
line by an integral is not sufficiently accurate for all $T$ (it
sets the coth factor equal to one). Setting the temperature dependent
result found in Eq.~(\ref{eq:fzero}) to $N_{b}/N$ from Eq.~(\ref{eq:N0integral}) gives us an equation for
$\gamma$, which can
be used to evaluate the individual energy-state occupation fractions
$n_{s_{x},0}/N$. The results are shown in Fig. \ref{figChannel}.

Near the condensation temperature the number of macroscopic states
involved is clearly very large. Only near $T=0$ does the lowest state
become dominant. All these occupation numbers are of order $N.$

\section{The Cigar Trap\label{sec:The-Cigar-Trap}}

For another, and most detailed, example of GBEC we consider the anisotropic
harmonic trap. A few years ago Van Druten and Ketterle (VDK) \cite{VDK}
suggested an unusual BEC that would take place in a cigar-shaped harmonic
trap (one long dimension with a weak harmonic potential, plus two
short dimensions with much stronger transverse potentials). More recently
Beau and Zagrebnov (BZ) \cite{BeauZag} independently treated the
same problem. In each treatment there is a GBEC into a band of states,
and a second transition at a lower temperature. In these treatments
we learn that there are two ways to take the thermodynamic limit (TL); in one the transition
disappears (that is, it is a pseudo-transition), while in the other
it persists in the infinitely large system. This peculiarity illustrates
some interesting unexpected features of the statistical mechanics
of BEC. The approach used by BZ to fix this second transition was
originally invented by Van den Berg and co-workers \cite{VDB83,VDB86}
for an anisotropic square well potential.

The energy levels of the cigar trap are \begin{equation}
\varepsilon_{p_{x}p_{y}p_{z}}=\hbar\omega_{\perp}(p_{x}+p_{y})+\hbar\omega_{\|}p_{z},\end{equation}
 where $\omega_{i}=\sqrt{U/m}/L_{i}$ corresponding to the two lengths
$L_{\perp}$ and $L_{\|}$, with $L_{\|}\gg L_{\perp}$. The density
parameter of the gas is then taken to be \begin{equation}
\rho=\frac{N}{L_{\perp}^{2}L_{\|}}.\label{eq:density}\end{equation}
 We again define a characteristic temperature $T_{0}$ just as in
Eq.~(\ref{eq:T0}). With this notation the particle number must satisfy
\begin{equation}
N=\sum_{p_{x},p_{y},p_{z}}\frac{1}{\exp\left[\frac{T_{0}a}{T}\left(\frac{p_{x}+p_{y}}{L_{\perp}}+\frac{p_{z}}{L_{\|}}\right)+\alpha\right]-1}.\end{equation}
Change the sum to an integral for $T$ large to give \begin{eqnarray}
N & = & \int_{0}^{\infty}dp_{x}\int_{0}^{\infty}dp_{y}\int_{0}^{\infty}dp_{z}\frac{1}{\exp\left[\frac{T_{0}a}{T}\left(\frac{p_{x}+p_{y}}{L_{\perp}}+\frac{p_{z}}{L_{\|}}\right)+\alpha\right]-1}\nonumber \\
 & = & \left(\frac{T}{T_{0}}\right)^{3}\frac{L_{\perp}^{2}L_{\|}}{a^{3}}\int_{0}^{\infty}du\int_{0}^{\infty}dv\int_{0}^{\infty}dw\frac{1}{\exp\left[u+v+w+\alpha\right]-1}\nonumber \\
 & = & \left(\frac{T}{T_{0}}\right)^{3}N\sum_{l=1}^{\infty}e^{-\alpha l}\int_{0}^{\infty}du\int_{0}^{\infty}dv\int_{0}^{\infty}dwe^{-l(u+v+w)}\nonumber \\
 & = & \left(\frac{T}{T_{0}}\right)^{3}N\sum_{l=1}^{\infty}\frac{e^{-\alpha l}}{l^{3}}=N\left(\frac{T}{T_{0}}\right)^{3}F_{3}(\alpha).\end{eqnarray}
 Thus the chemical potential parameter $\alpha$ has to satisfy\begin{equation}
\left(\frac{T}{T_{0}}\right)^{3}F_{3}(\alpha)=1.\end{equation}
If $T$ becomes too small, the equation can no longer be satisfied.
The condensation temperature is \begin{eqnarray}
T_{c} & = & \frac{T_{0}}{\zeta(3)^{1/3}}\nonumber \\
 & = & \frac{\hbar}{k_{B}}\sqrt{\frac{U}{m}}\left(\frac{\rho}{\zeta(3)}\right)^{1/3},\end{eqnarray}
 where the density $\rho$ is given by Eq.~(\ref{eq:density}).

The number of noncondensate particles now satisfies the relation\begin{equation}
N-N_{0}=N\left(\frac{T}{T_{c}}\right)^{3},\end{equation}
 with $N_{0}$ the condensate number or \begin{equation}
f_{0}(T)\equiv\frac{N_{0}}{N}=1-\left(\frac{T}{T_{c}}\right)^{3}.\label{eq:fb}\end{equation}
When the temperature is below $T_{c}$ the condensation is in a \emph{band}
of states, as we will show explicitly below. The band occupation number
is\begin{eqnarray}
N_{b} & = & \sum_{p_{z}=0}^{\infty}\frac{1}{\exp\left[\frac{T_{0}a}{T}\frac{p_{z}}{L_{\|}}+\alpha\right]-1}.\label{eq:Nb}\end{eqnarray}
 If we change the sum to an integral we find\begin{eqnarray}
N_{b} & = & \frac{TN}{T_{0}K}\int_{0}^{\infty}dw\frac{1}{\exp\left[w+\alpha\right]-1},\label{eq:NbIntegral}\end{eqnarray}
 in which we have made the substitution\[
\frac{L_{\|}}{a}=\left(\frac{N}{L_{\perp}^{2}L_{\|}}\right)^{1/3}L_{\|}=\frac{N}{K},\]
 where\begin{equation}
K\equiv\left(N\frac{L_{\perp}}{L_{\|}}\right)^{2/3}.\label{eq:K}\end{equation}
 Note the assumption (to be corrected later) in this replacement of
the sum by an integral, that the ground state occupation remains as
small as that of the rest of the states. Carrying out the integration
gives\begin{equation}
f_{b}\equiv\frac{N_{b}}{N}=-\frac{T}{T_{0}K}\ln\left(1-e^{-\alpha}\right).\end{equation}
 If indeed the ground-state occupation is negligible, then the band
occupation is the whole condensate, $f_{b}=f_{0}$ and, since $\alpha$
is very small below the transition temperature, we have $\ln(1-e^{-\alpha})\approx\ln\alpha$;
we can solve for $\alpha$ to find\begin{equation}
\alpha=e^{-f_{0}\frac{T_{0}}{T}K}.\label{eq:channelalpha}\end{equation}
The band states are analogous to a one-dimensional (1D) gas of Bose
particles in the long harmonic trap. Does this 1D gas also have its
own condensation into its lowest-energy state, which is, of course,
the overall ground state? Subsequent treatment now depends on the
boundary conditions used in taking the TL.

\subsection{Standard boundary conditions}

Under normal thermodynamic limit conditions, the ratio \begin{equation}
\Delta\equiv\frac{L_{\|}}{L_{\perp}}\end{equation}
is held constant, while $N,L_{\|,}$ and $L_{\perp}$ are increased.
In the present TL, $K=\left(N/\Delta\right)^{2/3}$ (Eq.~(\ref{eq:K}))
would be increasingly large as $N$ increases ($\sim N^{2/3})$ and,
for $T\lesssim T_{c}$, $\alpha$ would be increasingly small going
as $\alpha\sim e^{-N^{2/3}}$from Eq.~(\ref{eq:channelalpha}).

The occupation numbers of the low-lying band levels are\begin{equation}
n_{00p_{z}}\cong\frac{1}{\frac{T_{0}K}{T}\frac{p_{z}}{N}+\alpha}.\end{equation}
 The first term in the denominator being of order $N^{-1/3}$ dominates
$\alpha$ giving $n_{00p_{z}}=O(N^{1/3})$ and $n_{00p_{z}}/N\rightarrow0$
in the TL (microscopic occupation); the condensation is then of Type
III as listed in Sec. \ref{sec:Introduction}. The $p_{z}=0$ term
is the exception; when $n_{000}/N$ is microscopic, the condensation
remains of Type III. However, the ground state occupation can become
$O(N)$ when \begin{equation}
n_{000}\cong\frac{1}{\alpha}=e^{f_{0}\frac{T_{0}}{T}K}=cN,\end{equation}
 where $c\lesssim1.$ This occurs at a temperature $T_{1}$ given
by \begin{equation}
\frac{T_{1}}{f_{0}(T_{1})}=T_{0}\frac{K}{\ln cN},\label{eq:T1B}\end{equation}
 or, assuming $N$ large and $T_{1}$ is sufficiently smaller than
$T_{c}$ that $f_{0}\approx1,$ we have\begin{equation}
T_{1}=\frac{\hbar}{k_{B}}\sqrt{\frac{U}{m}}\frac{N}{L_{\|}\ln N},\end{equation}
 which might be considered the condensation temperature of the 1D
band. In the TL, $T_{1}$ would go to zero as $1/\ln N$ \emph{if}
the 1D density $N/L_{\|}$ were held constant. Such a behavior is
indeed characteristic of a 1D pseudo-phase-transition \cite{WJM1D3D,VDK1D3D},
which does not exist except in a finite system. While it is possible
to consider a 3D TL such that $L_{\perp}$ is held constant while
$N$ and $L_{\|}$ increase, it is $\rho=N/(L_{\perp}^{2}L_{\|})$
and $\Delta$ that are held constant in the standard 3D TL while the
two lengths increase. In Eq.~(\ref{eq:T1B}), when the right side
becomes very large, $\sim N^{2/3}/\ln N$, the only way the equation
can be satisfied is to have $T_{1}\approx T_{c}$ so that the denominator
on the left, $f_{0}(T_{1}),$ almost vanishes (See Eq.~(\ref{eq:fb})).
Thus in the TL, $T_{1}\rightarrow T_{c}$ and the {}``1D'' transition
coincides with the condensation into the band of states, that is,
the band then collapses to just the ground state and the system \emph{reverts
to the normal BEC} rather than a GBEC. Such a result is rather surprising
and has not been noted before in the literature. See below for a numerical
example.

When the ground state occupation becomes large, the analysis of Eq.
(\ref{eq:NbIntegral}) is no longer valid, and we need to be a bit
more careful to find an expression for the ground-state occupation
number. In the sum of Eq.~(\ref{eq:Nb}) we should split off the ground
state term $n_{g}$ to write\begin{eqnarray}
N_{0} & = & n_{g}+N_{b}\nonumber \\
 & = & n_{g}+\sum_{p_{z}^{\prime}=0}^{\infty}\frac{1}{\exp\left[\frac{T_{0}K}{T}\frac{p_{z}^{\prime}}{N}+\alpha^{\prime}\right]-1},\end{eqnarray}
 where we have redefined the summation index $p_{z}^{\prime}=p_{z}-1$
and let\begin{equation}
\alpha^{\prime}=\alpha+\frac{T_{0}K}{NT}.\end{equation}
 The ground-state occupation fraction is \begin{equation}
f_{g}=\frac{1}{N\left(e^{\alpha}-1\right)}.\label{eq:ng}\end{equation}
 Replace the sum by an integral as before to give\begin{equation}
f_{0}=\frac{1}{N}\frac{1}{e^{\alpha}-1}-\frac{T}{T_{0}K}\ln\left[1-e^{-\left(\frac{T_{0}K}{NT}+\alpha\right)}\right],\label{eq:fbrigorous}\end{equation}
 Since the total condensate $f_{0}$ is given by Eq.~(\ref{eq:fb}),
this is a transcendental equation for $\alpha.$

A formula to solve self-consistently for $f_{g}$ for all situations
is gotten from Eq.~(\ref{eq:fbrigorous}) by solving Eq.~(\ref{eq:ng})
for $e^{\alpha}$ in terms of $f_{g}$: \begin{equation}
f_{g}=f_{0}+\frac{T}{T_{0}K}\ln\left[1-e^{-\frac{T_{0}K}{NT}}\frac{1}{1+1/(Nf_{g})}\right].\label{eq:foSelfCon}\end{equation}
In the TL the factor in front of the logarithm in Eq.~(\ref{eq:foSelfCon})
dominates the behavior and the second term goes to zero so $f_{g}\rightarrow f_{0}$
as we stated above.

\subsection{Numerics}

\begin{figure}[h]
 \centering \includegraphics[width=3in]{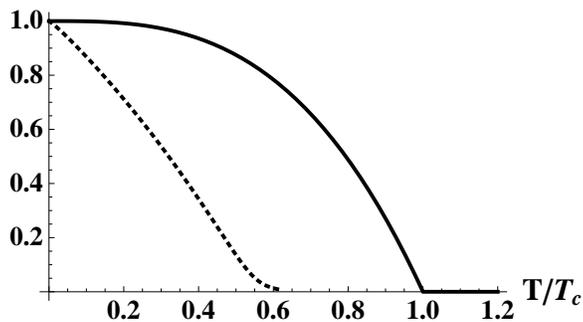}

\caption{The condensate distributions in the standard boundary limiting case
(the ratio of boundary lengths $\Delta=L_{\|}/L_{\perp}$is a fixed
constant) and for $N=10^{6},$ $\Delta=5.6\times10^{4}$. The solid
curve represents the full condensate $f_{0}$ (the fraction of particles
in the lowest band of states), while the dotted curve is the ground-state
pseudo-condensate $f_{g}$ (the fraction in the lowest state of the
lowest band). }

\label{fig2} 
\end{figure}

Generally we have \begin{equation}
\frac{T_{c}}{T_{0}}=\frac{1}{\zeta(3)^{1/3}}=0.94.\end{equation}
 VDK considered a set of parameters with $N=10^{6}$ and $\Delta=5.6\times10^{4}$.
In the this case we have \begin{eqnarray}
K & \equiv & \left(N\frac{L_{\perp}}{L_{\|}}\right)^{2/3}=\left(\frac{10^{6}}{5.6\times10^{4}}\right)^{2/3}=6.8,\\
\frac{T_{1}}{T_{c}} & = & f_{0}(T_{1})\frac{K\zeta(3)^{1/3}}{\ln N}=0.47\quad.\label{eq:T1}\end{eqnarray}
 The numerical result in Eq.~(\ref{eq:T1}) was gotten by iteration;
if we set $f_{0}=1$ in the $T_{1}$ formula we get $T_{1}/T_{c}=0.52$;
putting that back on the right in $f_{0}(T_{1})$ gives a new value
of $T_{1}$, etc. In Fig. \ref{fig2} we plot the results for $f_{0}$
and $f_{g}$, where we have solved Eq.~(\ref{eq:foSelfCon}) by simple
iteration.

\begin{figure}[h]
 \centering \includegraphics[width=3in]{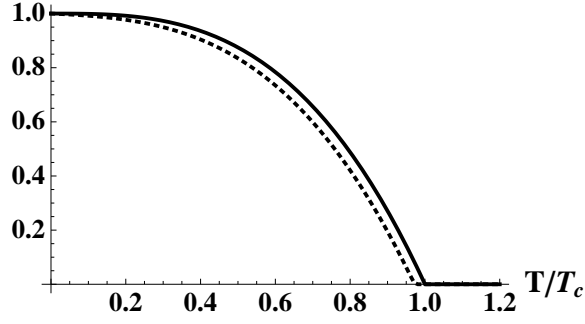}

\caption{The condensate distributions in the standard boundary limiting case
(the ratio of boundary lengths $\Delta=L_{\|}/L_{\perp}$is a fixed
constant) and for $N=10^{8},$ $\Delta=5.6\times10^{4}$. The solid
curve represents the full condensate $f_{0}$, while the dotted curve
is the ground-state pseudo-condensate $f_{g}$. }

\label{fig3} 
\end{figure}

In the case where we keep the \emph{same} length ratio, but increase
the number of particles to $N=10^{8}$ we have\begin{eqnarray}
K & = & \left(\frac{10^{8}}{5.6\times10^{4}}\right)^{2/3}=147,\\
\frac{T_{1}}{T_{c}} & = & \frac{f_{0}(T_{1})K\zeta(3)^{1/3}}{\ln N}=0.961\quad.\end{eqnarray}
 Fig. \ref{fig3} shows the plots of the two condensate fractions
for $N=10^{8}$. The ground-state transition almost coincides with
the band condensate as expected from the analytic argument.%

$T_{1}$ approaches $T_{c}$ because in Eq.~(\ref{eq:T1B}) the right
side becomes very large with $K\sim N^{2/3}\rightarrow\infty$, requiring
increasingly large $1/f_{0}(T_{1})$ on the left as explained above.
The question then arises whether there is a TL such that the right
side does \emph{not} become large, that is, in which $K\sim\ln N$.
In the next section we will find such a case.

\subsection{Exponential boundary conditions}

\begin{figure}[h]
 \centering \includegraphics[width=3in]{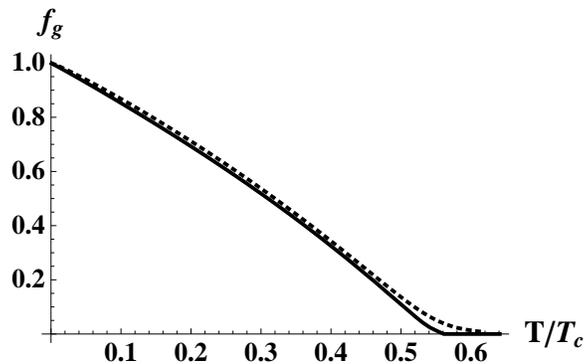}

\caption{The ground-state condensate distributions $f_{g}$ in the BZ boundary
limiting case (the ratio of boundary lengths $L_{\|}/L_{\perp}=e^{gL_{\perp}^{2}}$
is not a fixed constant in the thermodynamic limit). The dotted curve
is for $N=10^{6}$ while the solid curve is for $N=10^{16}$. }

\label{fig4} 
\end{figure}

\begin{figure}[h]
\centering \includegraphics[width=3in]{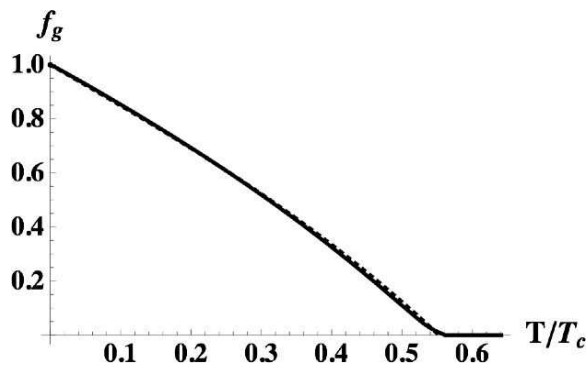}

\caption{The ground-state condensate distributions $f_{g}$ in the BZ boundary
limiting case for $N=10^{16}$ (solid curve) and the thermodynamic
limiting value, Eq.~(\ref{eq:Final}) (dotted curve). }

\label{fig5} 
\end{figure}

Here we consider how the above discussion is changed if we now take
the length ratio to obey \begin{equation}
\Delta=\frac{L_{\|}}{L_{\perp}}=e^{gL_{\perp}^{2}},\label{eq:Delta}\end{equation}
a boundary condition that certainly forces the system to become more
1D as $L_{\perp}$ increases. Similar exponential boundary relations
were first proposed by Van den Berg and co-workers \cite{VDB83,VDB86}
for an anisotropic square well potential. But these explicit conditions
for the anisotropic harmonic oscillator, in which the length factor
in the exponent is squared were given by Beau and Zagrebnov \cite{BeauZag,Beau}
and we designate them as BZ conditions. That they fit the requirement
that $K\sim\ln N$ will be verified below. It becomes useful to use
a unitless notation here. Let\begin{eqnarray}
\ell_{\perp} & = & \rho^{1/3}L_{\perp},\nonumber \\
\ell_{\|} & = & \rho^{1/3}L_{\|},\nonumber \\
\gamma & = & \frac{g}{\rho^{2/3}}.\end{eqnarray}
 Then, from the density relation Eq.~(\ref{eq:density}), we have
\begin{eqnarray}
N & = & \ell_{\perp}^{2}\ell_{\|}=\ell_{\perp}^{3}e^{\gamma\ell_{\perp}^{2}}=\ell_{\perp}^{3}\Delta,\nonumber \\
K & = & \ell_{\perp}^{2},\nonumber \\
\gamma & = & \frac{\ln\Delta}{\ell_{\perp}^{2}}.\label{eq:Nell}\end{eqnarray}
 The VDK parameters $N=10^{6}$ and $\Delta=5.6\times10^{4}$ can
be considered a special case (one particular $N$ value) of a BZ set.
Then $N$ and $\Delta$ give the corresponding $\gamma$ value. From
the above formulas we have $\ell_{\perp}=2.61$ and $\gamma=1.60,$
which is the $\gamma$ value we will use later for larger $N$ values
at the same density. For the original VDK parameters, the curves for
$f_{0}$ and $f_{g}$ are obviously identical to those shown in Fig.
\ref{fig2}; however, with the condition Eq.~(\ref{eq:Delta}) and
a larger $N$ value, $f_{g}$ will no longer merge with the $f_{0}$
as we will see. We consider a much larger $N$ value and solve Eq.
(\ref{eq:foSelfCon}) for $f_{g}$. We show the comparison of the
$f_{g}$ curves for $N=10^{6}$ and $N=10^{16}$ in Fig. \ref{fig4}.
We can understand the behavior of the large $N$ curve by approximating
Eq.~(\ref{eq:foSelfCon}): For large $N$ (any we have used here)
we can expand the exponential and the fraction inside the logarithm
to find \begin{equation}
f_{g}=f_{0}+\frac{T}{T_{0}K}\ln\left[\frac{T_{0}K}{NT}+\frac{1}{Nf_{g}}\right]\approx f_{0}-\frac{T}{T_{0}K}\ln\left[N\right],\label{eq:foSelfConApp}\end{equation}
 where the last approximation holds because $N$ is much larger than
the other factors in the logarithm. The parameter $K=\ell_{\perp}^{2}$
can be expressed approximately as well. From Eq.~(\ref{eq:Nell})
we write\begin{equation}
\gamma\ell_{\perp}^{2}=\ln N-\frac{3}{2}\ln\ell_{\perp}^{2}.\end{equation}
 By iterating this formula once, we see that the second term is much
smaller than the first and so\begin{equation}
K=\ell_{\perp}^{2}\approx\frac{\ln N}{\gamma}.\end{equation}
 Putting this in Eq.~(\ref{eq:foSelfConApp}) gives\begin{equation}
f_{g}(T)=f_{0}(T)-\frac{T\gamma}{T_{c}\zeta(3)^{1/3}}.\label{eq:Final}\end{equation}
 We plot this result in comparison with the $N=10^{16}$ result in
Fig. \ref{fig5}. %

The lower transition temperature is gotten by setting the left side
of Eq.~(\ref{eq:Final}) to zero: \begin{equation}
T_{1}=\frac{T_{c}f_{0}(T_{1})\zeta(3)^{1/3}}{\gamma},\end{equation}
which must be solved self-consistently. We find $T_{1}=0.552T_{c}.$
Clearly this approximation is very good and the system has two distinct
phase transitions as BZ have claimed. Perhaps we should not be too
surprised that this was possible. In the VDK case having $T_{1}\thickapprox0.5T_{c}$
was arranged by a judicious choice of $\Delta$ for $N=10^{6}$. The
question then is whether it is possible, for a larger value of $N,$
to find a $\Delta$ such that $T_{1}$ and $\rho$ are both unchanged.
Such a value is given by Eq.~(\ref{eq:Delta}). VDK also numerically
treated a case in which $T_{1}$ was fixed while $N$ increased, although
they did not specify how this was done; they apparently used a form
equivalent to Eq.~(\ref{eq:Delta}).

There is a case of box boundary conditions (square-well potential)
that is isomorphic to the harmonic system treated in this section.
This is the flat plate geometry in which two large square plates of
length $L$ on a side are separated by a much smaller distance $D.$
This problem was treated by Van den Berg and coworkers \cite{VDB83,VDB86}
in the 1980's. A third analogous case involves having two free dimensions
and one harmonic dimension \cite{BeauZag}.

Sonin \cite{Sonin} gave an analysis of multi-step quasi-condensations in finite 
systems  for anisotropic free-particle boundary conditions 
and showed that a second transition could be preserved 
in the thermodynamic limit by an appropriate thermal limiting procedure.
Deng \cite{Deng}  and Shiokawa \cite{Shiokawa}  treated multi-step transitions in 
finite anisotropic harmonic potentials;  the extra transitions disappear in the
normal TL.

\section{GBEC in a box: The Casimir prism\label{sec:Type-III-GBEC-Casmir}}

\begin{figure}[h]
 \centering \includegraphics[width=3in]{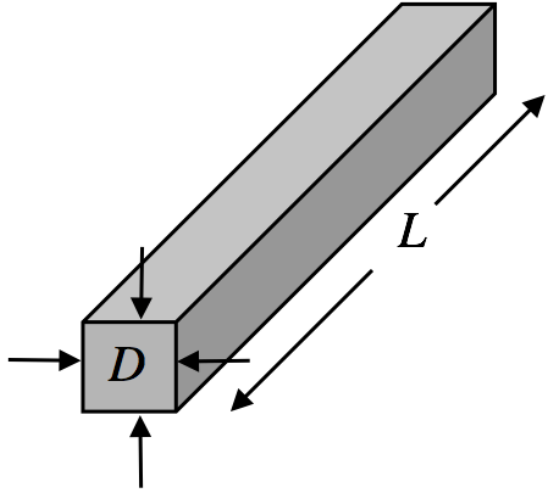}

\caption{ Casimir Prism. A box with one side $L$ much greater than the square
end face dimension $D.$}

\label{figCasmir} 
\end{figure}

We consider one more Type III GBEC where there is again a macroscopic
condensation into a band number of states while the occupation of
any single quantum state remains microscopic. With this geometry there
is no second transition at a lower temperature. The case considered
here was treated by Casimir \cite{Casimir} and later by others \cite{VDB82A,VDB86,Zag2004,BeauZag,Beau}.
It was one of the first known theoretical cases of GBEC. The geometry
is shown in Fig. \ref{figCasmir}. The length $L$ is much larger
than the side $D$ of the square cross section. %

The free-particle states in this system are given by\begin{equation}
\varepsilon_{s_{x}s_{y}s_{z}}=\frac{h^{2}}{2m}\left[\frac{(s_{x}^{2}+s_{y}^{2})}{D^{2}}+\frac{s_{z}^{2}}{L^{2}}\right],\end{equation}
 where the $s_{i}$ are again positive or negative integers. The total
number of particles is a sum \begin{equation}
N=\sum_{s_{x}s_{y}s_{z}}\frac{1}{\exp\left\{ \beta\left[\frac{h^{2}}{2m}\left(\frac{(s_{x}^{2}+s_{y}^{2})}{D^{2}}+\frac{s_{z}^{2}}{L^{2}}\right)-\mu\right]\right\} -1}.\end{equation}
 Above any transition we can change the sum to an integral:\begin{eqnarray}
N & = & \int_{-\infty}^{\infty}ds_{x}\int_{-\infty}^{\infty}ds_{y}\int_{-\infty}^{\infty}ds_{z}\frac{1}{e^{\beta\left[\frac{h^{2}}{2m}\left(\frac{(s_{x}^{2}+s_{y}^{2})}{D^{2}}+\frac{s_{z}^{2}}{L^{2}}\right)-\mu\right]}-1}\nonumber \\
 & = & \frac{LD^{2}}{h^{3}}\left(\frac{2m}{\beta}\right)^{3/2}\int_{-\infty}^{\infty}du_{1}\int_{-\infty}^{\infty}du_{2}\int_{-\infty}^{\infty}du_{3}\frac{1}{e^{\left(u_{1}^{2}+u_{2}^{2}+u_{3}^{2}+\alpha\right)}-1}\nonumber \\
 & = & \left(\frac{2\pi m}{\beta h^{2}}\right)^{3/2}LD^{2}F_{3/2}(\alpha).\label{eq:aboveTc}\end{eqnarray}
 $F_{3/2}(\alpha)$ has a maximum of $\zeta(3/2)$ at $\alpha=0$
so the transition temperature is given by\begin{equation}
T_{c}=\frac{h^{2}}{2\pi mk_{B}}\left(\frac{\rho}{\zeta(3/2)}\right)^{2/3}.\end{equation}
 For $T<T_{c}$, Eq.~(\ref{eq:aboveTc}) is no longer valid, and below
$T_{c}$ the condensed particle number $N_{0}$ satisfies \begin{equation}
N_{0}=N\left[1-\left(\frac{T}{T_{c}}\right)^{3/2}\right].\end{equation}
 Because of the anisotropy of the boundary conditions, a band of states
with $s_{x}=0,$ $s{}_{y}=0$ fills at the transition. To see how
this occurs, we examine the density of low excited states. There are
various ways to take the thermodynamic limit, but here we hold $D$
constant while we take $L\rightarrow\infty$. (The arguments presented
here also hold if we let $L,D\rightarrow\infty$ with $L$ approaching
infinity faster than $D^{2}.$ Compare with Appendix A.3.) Let us
first hypothesize that the ground state is macroscopically occupied
so that $\alpha\sim O(1/(D^{2}L))$ (which we will see is incorrect);
the occupation of the low states would then be\begin{eqnarray}
n_{s_{x},s_{y},s_{z}} & \approx & \frac{1}{\left[\frac{h^{2}\beta}{2m}\left(\frac{(s_{x}^{2}+s_{y}^{2})}{D^{2}}+\frac{s_{z}^{2}}{L^{2}}\right)+\alpha\right]}.\end{eqnarray}
 Under our hypothesis, the first term would dominate the other two,
which could be dropped and the state density would be $O(1)$ unless
$s_{x}=s_{y}=0.$ In that case, the $s_{z}$ term would still be negligible
and we would find $n_{0,0,s_{z}}\sim O(N)$ for \emph{all} values
of $s_{z}$. Clearly we need to sum up the entire band of $s_{z}$
states to find the true behavior. The number of particles in the band
is

\begin{eqnarray}
N_{b} & = & \sum_{s_{z}}\frac{1}{\exp\left[\frac{\beta h^{2}}{2m}\frac{s_{z}^{2}}{L^{2}}+\alpha\right]-1}\nonumber \\
 & = & \left(\frac{2m}{\beta h^{2}}\right)^{1/2}L\int_{-\infty}^{\infty}dw\frac{1}{\exp\left\{ w^{2}+\alpha\right\} -1}\nonumber \\
 & = & \left(\frac{2\pi m}{\beta h^{2}}\right)^{1/2}LF_{1/2}(\alpha).\end{eqnarray}
 Below the transition temperature, $\alpha$ is small and $N_{b}$
equals the condensate number of particles $N_{0}$. Further, since
\cite{Robinson} for small $\alpha$, $F_{1/2}(\alpha)\approx\sqrt{\pi/\alpha}$,
one gets\begin{equation}
N_{0}=\left(\frac{2\pi^{2}m}{\beta h^{2}}\right)^{1/2}\frac{L}{\sqrt{\alpha}}\end{equation}
 or\begin{eqnarray}
\alpha & = & \frac{2\pi^{2}m}{\beta h^{2}}\left(\frac{L}{N_{0}}\right)^{2}\nonumber \\
 & = & \frac{2\pi^{2}m}{\beta h^{2}}\frac{1}{\rho_{0}^{2}D^{4}},\label{eq:Casmiralpha}\end{eqnarray}
 where $\rho_{0}=N_{0}/LD^{2}$. This result is quite different from
what it would have been if the ground state were macroscopically occupied
as in our initial hypothesis of $\alpha\sim1/(D^{2}L)$.

The condensate number is macroscopic, but the numbers in the single-particle
states are not. The density corresponding to low quantum numbers $\left\{ 0,0,s_{z}\right\} $
is\begin{eqnarray}
\rho_{0,0,s_{z}} & \approx & \frac{1}{LD^{2}\left[\frac{h^{2}\beta}{2m}\frac{s_{z}^{2}}{L^{2}}+\frac{\gamma(T)}{D^{4}}\right]}\nonumber \\
 & = & \frac{1}{\left[\frac{h^{2}\beta}{2m}s_{z}^{2}\frac{D^{2}}{L}+\frac{\gamma L}{D^{2}}\right]},\label{eq:Expandedden}\end{eqnarray}
 where $\gamma=\alpha D^{4}$. The term in $\gamma$ always dominates
so the occupation of each of these states is microscopic and approaches
zero as $1/L$, \emph{including} the ground state with $s_{z}=0$.
Of course, at $T=0$ the ground state must finally have all the particles
in it, but the temperature at which this occurs can be estimated by
using $\alpha$ from Eq.~(\ref{eq:Casmiralpha}) in $\rho_{0,0,0}\approx1/(N\alpha)$
to see that the onset temperature for macroscopic occupation of the
ground state is $T\sim T_{c}$$\rho^{1/3}D^{2}/L$, which is zero
in the TL or very small in a real experiment.

There is an alternative way to take the limit in which all three dimensions
of the box become infinite. This way can also be used to distinguish
the three kinds of GBEC. We discuss this in the Appendix.

\section{The effects of interactions on GBEC\label{sec:interactions}}

\subsection{GBEC and interactions in the literature}

The purpose of this paper has been to outline the possible kinds of
GBEC by using ideal gases. It is not our purpose to make a complete
analysis of the existence of GBEC with arbitrary interactions. Nevertheless
it makes sense to ask whether GBEC would disappear with interactions.
Noziéres \cite{Nozieres2} has shown, in the Hartree-Fock approximation
in a homogeneoous scalar Bose gas that repulsive exchange interactions
favor condensation into a single state. He ignored the case of attractive
interactions, because they would cause the homogeneous system to collapse.
More importantly, he also assumed that each condensate state is \emph{macroscopically}
occupied, which applies only to Type I or II condensation. However,
in a trap a small degree of attraction does \emph{not} necessarily
lead to collapse \cite{BaymPethick}; the kinetic energy stablizes
the system. The literature also contains examples of GBEC's in interacting
systems \cite{Zag98-1Int,Zag98-2Int,Zag98-3Int,Zag2000Int,Zag2000-1Int,Schroder90Int,MichVerbeureInt},
involving repulsions and Type III band occupation. Girardeau \cite{Girar3}
considered an attractive interaction in a uniform system, which showed
macroscopic occupation of each condensate state, but he did not take
into account the possible collapse of the system. 

A common type of interacting model showing GBEC has \emph{diagonal}
interactions. These interactions are a function only of the number
of particles $N_{k}$ in the $k$th momentum state. Then the Hamiltonian
is a function of a set of mutually commuting operators with a particularly
simple spectrum. (See for example, Ref.~\cite{Zag98-1Int} and references
therein.) While one finds Type III GBEC in the interacting case, interactions
also introduce yet another type of BEC called \emph{non-convential}
or \emph{dynamic }condensation. The conventianal condensation occurs
when there is a kind of saturation: the total particle number becomes
larger than some critical value as in the NBEC or even Type III. A
dynamic condensation occurs only when induced by attractive interactions.

\subsection{The Hohenberg theorem }

We treated only one case of a 2D system, that of Sec. \ref{sec:The-Canal}
where we found a Type II condensation, that is, having a band of macroscopically
occupied states. However, the Hohenberg theorem \cite{Hohenberg} states that no macroscopic
condensation can occur into a zero-momentum state in two dimensions
in the thermodynamic limit. Since the theorem refers to condensation
into a \emph{momentum} state it might not seem to apply to the condensation
discussed in Sec. \ref{sec:The-Canal}, since one dimension, at least,
involves a harmonic potential. However, a theorem developed by Chester
\cite{Chester} based on work by Penrose and Onsager \cite{PenroseOnsager}
states that there can be \emph{no condensation into }any\emph{ state
unless there is one into the $k=0$ state}. The loophole relative
to the trapped gas is the condition assumed by the Chester derivation
that the density be finite everywhere in the thermodynamic limit.
As we show below, the gas studied in Sec. \ref{sec:The-Canal} has
a divergent density at the origin in the TL. If, however, the system
has repulsive interactions, no such divergence would be allowed and
then the Hohenberg theorem would apply and the transition studied
in that section would disappear.  A similar situation has
been discussed for a rotating Bose gas \cite{RehrMermin} and with
the 2D completely trapped gas \cite{WJM1D3D}.

To see the divergence at the origin for the system in Sec. \ref{sec:The-Canal}
we consider just the ground state contribution to the density. If
$\psi_{0}(z)$ is the ground-state harmonic wave function, then \begin{eqnarray}
\rho_{0}(0) & = & \frac{1}{L}\left|\psi_{0}(0)\right|^{2}n_{00}=\left(\frac{m\omega}{\pi\hbar}\right)^{1/2}\frac{n_{00}}{L}\nonumber \\
 & = & \left(\frac{\sqrt{mU}}{\pi\hbar}\right)^{1/2}\frac{n_{00}}{L^{3/2}}=O(N^{1/4}),\end{eqnarray}
 since $n_{00}=O(N)$ and $L=\sqrt{N/\rho}$. In the TL, the density
diverges, and the Chester theorem does not apply.

The next question then is whether a Hohenberg-like theorem forbids
a 2D transition of Type III, in which a band of \emph{microscopically}
occupied states condenses. This subject has been addressed \cite{WJMHohenTheorem}
and the usual derivation was shown \emph{not} to forbid such a transition.
However, the derivation in this reference does not tell us whether
some other theoretical approach might not reveal such an alternative
theorem forbidding the 2D transition. The question of whether a Type
III GBEC is possible in 2D seems still an open question. Recently
an analysis \cite{Fernadez} of an interacting 2D trapped gas showed
no fragmentation. In 1D an analysis similar to that in Ref.~\cite{WJMHohenTheorem}
shows that no GBEC can occur in the standard type of thermodynamic
limit.

\subsection{The 3D transition to 1D }

One of our prime examples of GBEC in Sec.~\ref{sec:The-Cigar-Trap}
involved the crossover between a 3D and a 1D gas in a cigar trap.  The
literature on bosons in cigar traps is much too large to summarize
here and no definitive GBEC has yet been seen experimentally.  A good
review is the paper of Bouchoule et al \cite{BouchouleVD} and a recent
relevant experiment is described by Armijo et al
\cite{ArmijoBouchoule}.  Actual experiments here are on finite
systems, of course, and sharp transitions are not observed.  One
theoretical advantage of 1D is that there is an exact analytic
solution due to Lieb and Liniger \cite{LL} for bosons interacting by
a $\delta$-function interatom potential, with thermodynamics by Yang
and Yang \cite{YangYang}.  This theory has been used extensively in
analyses of the experiments.  Forrester et al \cite{Forrester} have
been able to show that, for the 1D Bose gas with \emph{infinite}
$\delta$-function interaction, whether homogeneous or harmonically
trapped, all the eigenvalues of the one-body density matrix at $T=0$
are of order $\sqrt{N}$.  Such a result would certainly correspond to
a GBEC if we could state that the system was a quantum fluid.  But
this impenetrable gas would seem more like a solid than a fluid!

Armijo et al \cite{ArmijoBouchoule} map out the dimensional crossover
from a 3D gas to a 1D gas in a cigar trap. They observe a transition,
which scales experimentally exactly according to the description in
which {}``atoms accumulate in the transverse ground state, although
no single quantum state is macroscopically occupied.'' This is precisely
what we describe as GBEC in Sec.~\ref{sec:The-Cigar-Trap}. Theoretically
no lower sharp second transition to a true BEC, in which only the
ground state is occupied is expected, but rather a ``quasicondensate''
is formed. The formation of the quasicondensate is driven by interactions,
which surpress density fluctuations while the phase still fluctuates
\cite{Petrov}. The observations conform to this description. 

In this regard the path-integral-Monte-Carlo (PIMC) calculation of
Nho and Blume \cite{NhoBlume} on the 1D-3D crossover is particularly
relevant. They compute the superfluid component in the gas. With the
PIMC approach it is difficult to compute the occupation numbers for
the interacting gas. However, using those of the noninteracting gas,
they find that the superfluid component tracks the occupation of the
entire band corresponding to the lowest transverse harmonic state
much more accurately than the occupation of the lowest level in that
band. 

A further indication is the work of Witkowska et al \cite{Rzaz} in
which the authors use a {}``classical field approximation'' to study
the evaporative cooling dynamics of a trapped interacting 1D Bose
system. They compute the eigenvalues of the one-body density matrix
and find, in an intermediate temperature range, that the lowest four
states have occupations of $\sim10\%$ or more. As the temperature
goes even lower only the ground state remains occupied. This looks
much like a GBEC or at least a ``quasi-generalized-condensate.''

\subsection{Spin-1 Bose gas}

Ho and Yip \cite{Ho} have described a spin-1 Bose gas with antiferromagnetic
spin-spin interaction, which has a fragmented condensate ground state.
Here angular momentum conservation prevents spin flips between $+1$
and $-1$ states. The ground state is analogous to having a scaler
Bose gas in a double-well potential. A magnetic field gradient would
allow such spin transitions and results in a non-fragmented condensate
state.

\section{Conclusion}

Our purpose in this paper has been to give a tutorial on generalized
Bose-Einstein condensation to clarify what seems to be confusing literature
on the subject. The research in the field has gone in directions with
some researchers having been quite unaware of the results found by
others. Proofs that GBEC cannot exist, which, to some readers seemed
general, simply do not apply to other forms of the phenomenon of which
the authors were not even aware. 

We have seen that there are three types of GBEC and have given examples
of each. We have examined a form particular relevant to recent experiments,
that in the cigar trap, developed independently in Refs.~\cite{BeauZag}
and \cite{VDK} and have seen how the properties in the thermodynamic
limit can be completely different depending on how that limit is taken;
under a rather pecular limit there can be a two-stage condensation
in the ideal gas. It is possible that experiments in cigar traps have
already observed the upper transition, although it is unlikely that
there could be any sharp lower transition because of interactions
and because no thermodynamic limit is actually taken. 

Our hope is that this paper might stimulate further research in this
area, especially in interacting systems, so that ultimate experimental
verification of the existence of the phenomen might be observed.

\section*{Appendix: Alternative set of boundary conditions}

Van den Berg and co-workers \cite{VDB86} have used a uniform approach
for general boundary conditions that allows one to distinguish three
geometries under which the three kinds of BEC occur. However, these
geometries are \emph{not} in every case identical to the situations
we have discussed above as we will see. Consider a general box of
sides $L_{1},$$L_{2},$ and $L_{3}$; all three of these dimensions
will be taken to infinity in the thermodynamic limit. The TL involves
defining an arbitrary length parameter of, say, atomic size (we choose
the interparticle separation $a$) and a unitless parameter $H$ that
will be taken to infinity to establish the thermodynamic limit. We
define\begin{equation}
L_{i}=aH^{\nu_{i}},\end{equation}
 where the $\nu_{i}$ are fixed parameters such that the volume $V$
of the box is linear in $H:$\begin{equation}
V=L_{1}L_{2}L_{3}=a^{3}H,\end{equation}
 so that \begin{equation}
\nu_{1}+\nu_{2}+\nu_{3}=1.\end{equation}
The parameters are arranged according to \begin{equation}
\nu_{1}\ge\nu_{2}\ge\nu_{3}>0,\end{equation}
 and the three types of GBEC are categorized according to whether
$\nu_{1}$ is smaller than 1/2, equal to 1/2, or greater than 1/2.

We will assume we are below the 3D transition temperature. The density
of the lowest states is given by\begin{eqnarray}
\rho_{s_{1},s_{2},s_{3}} & = & \frac{n_{s_{1},s_{2},s_{3}}}{V}=\frac{1}{a^{3}H\left\{ \exp\left[\frac{\beta h^{2}}{2ma^{2}}\left(\frac{s_{1}^{2}}{H^{2\nu_{1}}}+\frac{s_{2}^{2}}{H^{2\nu_{2}}}+\frac{s_{3}^{2}}{H^{2\nu_{3}}}\right)+\alpha\right]-1\right\} }\nonumber \\
 & \approx & \frac{1}{a^{3}\left[\frac{h^{2}\beta}{2ma^{2}}\left(H^{1-2\nu_{1}}s_{1}^{2}+H^{1-2\nu_{2}}s_{2}^{2}+H^{1-2\nu_{3}}s_{3}^{2}\right)+\gamma\right]},\label{eq:statedensity}\end{eqnarray}
 where \begin{equation}
\gamma\equiv\alpha\frac{V}{a^{3}}.\end{equation}
 In every case the condensate density satisfies the usual 3D behavior,\begin{equation}
\rho_{0}=\rho\left[1-\left(\frac{T}{T_{c}}\right)^{3/2}\right],\end{equation}
 as shown in, say, Sec. \ref{sec:Type-III-GBEC-Casmir}.

\subsection*{A.1 Type I. $\nu_{1}<1/2$}

The NBEC in a cubic box corresponds to $\nu_{1}=\nu_{2}=\nu_{3}=1/3.$
More generally we see that for $\nu_{1}<1/2$ every $1-2\nu_{i}>0.$
Assume the smallest of these is $1-2\nu_{1}$. We hypothesize that
the ground state number is macroscopic so its density satisfies $a^{3}\rho_{000}=1/\gamma=O(1)$.
Then in Eq.~(\ref{eq:statedensity}) the terms in $s$ dominate over
the $\gamma$ term and an excited-state density is $\sim1/$$H^{1-2\nu_{1}}$
(if $s_{1}=0$, then $\rho_{0,s_{2},s_{3}}$ is $O(1/$$H^{1-2\nu_{2}})$
or $O(1/$$H^{1-2\nu_{3}})$, which is even smaller). In the thermodynamic
limit we have $H\rightarrow\infty$ and the excited state densities
vanish. Thus only the ground state is occupied and our hypothesis
is verified.

\subsection*{A.2 Type II. $\nu_{1}=1/2$}

In this case $1-2\nu_{1}=0.$ We again suppose that $\gamma=O(1),$
so that the dominant term in the square bracket of Eq.~(\ref{eq:statedensity})
is either that in $s_{2}$ or $s_{3}$ of order H$^{1-2\nu_{2,3}}$
so that the corresponding density vanishes as $H\rightarrow\infty.$
However, if $s_{2}=s_{3}=0$, then the term in $s_{1}$ is the same
order as $\gamma$ , so that there is macroscopic occupation of each
state, and we need to consider the whole $s_{2}=s_{3}=0$ band of
states. This band has density\begin{equation}
\rho_{band}=\sum_{s_{1}=-\infty}^{\infty}\frac{1}{a^{3}\left[\frac{h^{2}\beta}{2ma^{2}}s_{1}^{2}+\gamma\right]}.\label{eq:bandnu}\end{equation}
 In the present geometry it is no longer valid to replace the sum
in Eq.~(\ref{eq:bandnu}) by an integral since both terms in the denominator
are of the same order of magnitude. Here we must solve Eq.~(\ref{eq:bandnu})
for $\gamma.$ The sum can be done analytically in terms of a hyperbolic
cotangent (Sec. \ref{sec:The-Canal}) giving a transcendental equation
for $\gamma$.

\subsection*{A.3 Type III. $\nu_{1}>1/2$}

Now we have $1-2\nu_{1}<0$ while the other two such factors are positive.
The state density can be rewritten \begin{eqnarray}
\rho_{s_{1},s_{2},s_{3}} & \approx & \frac{1}{a^{3}\left[\frac{h^{2}\beta}{2ma^{2}}\left(s_{1}^{2}/H^{2\nu_{1}-1}+H^{1-2\nu_{2}}s_{2}^{2}+H^{1-2\nu_{3}}s_{3}^{2}\right)+\gamma\right]},\end{eqnarray}
 where each of the exponents is positive. In Type III all states are
microscopically occupied so we should try assuming that $\gamma=H^{\eta}$
where $\eta$ is a positive number less than $1-2\nu_{3}$. In that
case the order of magnitude of the densities with $s_{1}\neq0$ is
either $1/H^{1-2\nu_{3}}$ or $1/H^{1-2\nu_{2}}$, the latter case
occurring only if $s_{3}=0$. These densities vanish faster than the
densities of the band of states with $s_{2}=s_{3}=0$, $s_{1}\neq0$;
the latter are all of the same order of magnitude $1/a^{3}\gamma$
as the ground state and we need to sum these band states to get the
entire condensate:\begin{eqnarray}
\rho_{0} & = & \sum_{s_{1}=-\infty}^{\infty}\frac{1}{a^{3}\left[\frac{h^{2}\beta}{2ma^{2}H^{2\nu_{1}-1}}s_{1}^{2}+\gamma\right]}\nonumber \\
 & = & \frac{1}{a^{3}}\left(\frac{2ma^{2}H^{2\nu_{1}-1}}{h^{2}\beta}\right)^{1/2}\frac{\pi}{\sqrt{\gamma}}\coth\left(\pi\sqrt{\frac{2ma^{2}H^{2\nu_{1}-1}\gamma}{h^{2}\beta}}\right)\label{eq:TypeIII}\end{eqnarray}
 which yields\begin{equation}
\gamma=O(H^{2\nu_{1}-1}),\end{equation}
 so $\eta=2\nu_{1}-1$. In the limit of $H\rightarrow\infty$, $\rho_{s_{1},0,0}=O(H^{1-2\nu_{1}})\rightarrow0$
so each condensate state is microscopically occupied as assumed.

In Eq.~(\ref{eq:TypeIII}) the distribution is $\sim\left(\frac{\hbar^{2}\beta k^{2}}{2m}+\gamma\right)^{-1}$,
which will become negligibly small for momenta at $k>k_{0}$ with
\begin{equation}
\frac{\hbar^{2}\beta k_{0}^{2}}{2m}=C\gamma\end{equation}
 where $C$ is, say, $10^{4}$. The cutoff momentum is \[
k_{0}\sim1/H^{1-\nu_{1}}\rightarrow0\]
 in the TL; the condensate bandwidth in momentum space is vanishingly
small. Indeed Girardeau defines GBEC in the following way for a homogeneous
system \cite{Girar1}: He writes \begin{equation}
f=\lim_{k_{0}\rightarrow0}\lim_{N\rightarrow\infty}\frac{1}{N}\sum_{k<k_{0}}n_{k},\end{equation}
 defining {}``$f$ as the fraction of the total number of particles
with momenta less than any macroscopic momentum'' and where $n_{k}$
is the number of particles in momentum state $k$.

However, the quantum number corresponding to this cutoff momentum
is \begin{equation}
s_{0}=\frac{ak_{0}}{2\pi}H^{\nu_{1}}\sim H^{2\nu_{1}-1}\rightarrow\infty,\end{equation}
 since $\nu_{1}>1/2.$ So while the condensate band has all its significant
momenta microscopic in the TL, the number of such states in the condensate
is infinite!

\end{document}